Graphical Abstract

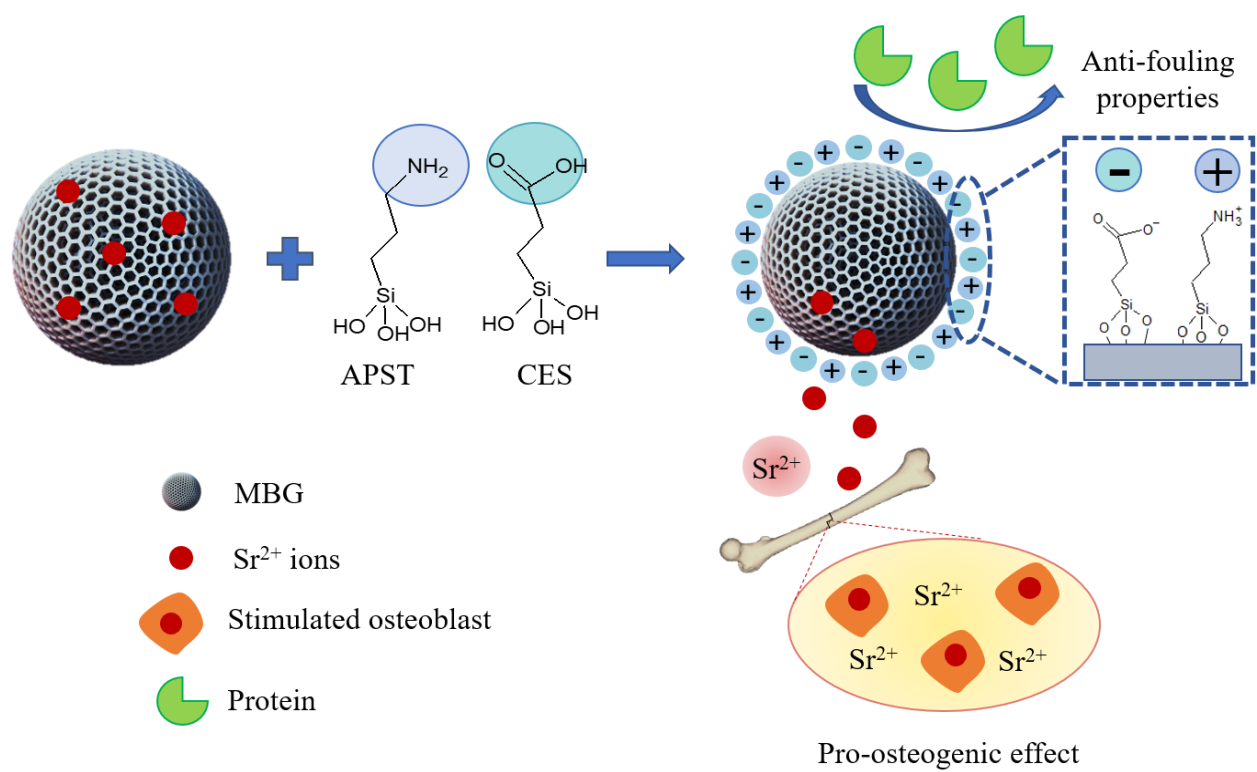

# Strontium-releasing mesoporous bioactive glasses with low-fouling *zwitterionic* surface as advanced biomaterials for bone tissue regeneration

Carlotta Pontremoli[1], Isabel Izquierdo-Barba[2,3], Giorgia Montalbano[1], María Vallet-Regí[2,3], Chiara Vitale-Brovarone[1], Sonia Fiorilli[1]*

1. Department of Applied Science and Technology, Politecnico di Torino, Corso Duca degli Abruzzi 24, Torino, Italy; carlotta.pontremoli@polito.it (C.P.); giorgia.montalbano@polito.it (G.M.); sonia.fiorilli@polito.it (S.F.); chiara.vitale@polito.it (C.V-B.)

2. Departamento de Química en Ciencias Farmacéuticas, Unidad de Química Inorgánica y Bioinorgánica, Universidad Complutense de Madrid. Instituto de Investigación Sanitaria Hospital 12 de Octubre i+12. Plaza Ramón y Cajal s/n, 28040 Madrid, Spain; : ibarba@ucm.es (I.I.B.); vallet@ucm.es (M.V.R.)

3. CIBER de Bioingeniería, Biomateriales y Nanomedicina, CIBER-BBN, Av. Monforte de Lemos, 3-5, 28029 Madrid, Spain.

*Correspondence: sonia.fiorilli@polito.it

**Abstract**

*Hypothesis*

The treatment of bone fractures still represents a challenging clinical issue when complications due to impaired bone remodelling (i.e. osteoporosis) or infections occur. These clinical needs still require a radical improvement of the existing therapeutic approach through the design of advanced biomaterials combining the ability to promote bone regeneration with anti-fouling/anti-adhesive properties able to minimise unspecific biomolecules adsorption and bacterial adhesion. Strontium-containing mesoporous bioactive glasses (Sr-MBG), able to exert a pro-osteogenic effect by releasing $Sr^{2+}$ ions, have been successfully functionalised to provide mixed-charge ($-NH_3^{\oplus}/-COO^{\ominus}$) surface groups with low-fouling abilities.

*Experiments*

Sr-MBG have been post-synthesis modified by co-grafting hydrolysable short chain silanes containing amino (aminopropylsilanetriol) and carboxylate (carboxyethylsilanetriol) moieties to achieve a *zwitterionic* zero-charge surface and then characterised in terms of textural-structural properties, bioactivity, cytotoxicity, pro-osteogenic and low-fouling capabilities.

*Findings*

After *zwitterionization* the *in vitro* bioactivity is maintained, as well as the ability to release $Sr^{2+}$ ions capable to induce a mineralization process. Irrespective of their size, Sr-MBG particles did not exhibit any cytotoxicity in pre-osteoblastic MC3T3-E1 up to the concentration of 75 µg/mL. Finally, the *zwitterionic* Sr-MBGs show a significant reduction of serum protein adhesion with respect to pristine ones. These results open promising future expectations in the design of nanosystems combining pro-osteogenic and anti-adhesive properties.

# 1. Introduction

Under normal healing conditions bony tissue completely regenerates in 6-8 weeks after injuries [1]. Despite this impressive regenerative ability, still up to 10–15% of the fracture patients show an impaired healing process, leading to a delayed healing outcome or even to a non-union [2]. Due to the increased of aging population, the total number of patients suffering from delayed bone healing or non-unions is expected to dramatically increase since elderly patients have higher risk of suffering from bone diseases (i.e. osteoporosis) and consequently for a compromised healing condition. Although the promising advancements in the design of novel biomaterials and therapeutic approaches to treat bone fractures and to support bone regeneration in a diseased environment, impaired healing outcomes are still a highly relevant clinical issue [3].

The underlying causes for an unsuccessful healing are various and depend among others on the facture site and severity, patient-dependent factors (e.g., age, chronic or autoimmune diseases) and associated complications, such as the occurrence of bacterial infections. In fact, bone infections remain one of the most potentially devastating impediments for bone healing, with important clinical and socio-economic implications [4–6]. Conventional treatments, based on systemic antibiotic administration [7], implant replacement and surgery[8], could result ineffective and lead to serious complications, causing limitations and significant reduction of the quality of patients' life such as serious side effects [9] and prolonged hospitalization [10]. In this regard, one of the main clinical concerns and causes for the failure of conventional treatments is the evolution and persistence of infections due to the ability of bacteria to form a biofilm [11,12], a microbial community characterized by cell embedded in a matrix of an own produced extracellular polymeric substances [13,14], which allow the bacteria becoming resistant to antimicrobial agents [14,15].

To overcome the clinical challenges associated to compromised bone regeneration, during the last years a plethora of nano-biomaterials have been proposed by the scientific community and used for the development of alternative therapeutic treatments [13–15]. Among the different biomaterials for bone tissue regeneration, mesoporous bioactive glasses (MBGs) are appealing candidates to develop multifunctional biomedical devices, due to their ability to synergistically combine the release of therapeutic ion/drug with anti-fouling/antibacterial properties. In fact, their extremely high exposed surface area and pore volume allow to store and release drugs (such as anti-inflammatory or antimicrobial agents) and biomolecules. Moreover, the composition of MBGs can be enriched through the incorporation of specific elements (i.e. Sr, Cu) with the aim to combine in a single biomaterial several therapeutic abilities, such as pro-osteogenic, pro-angiogenic and antibacterial properties. This promising and versatile approach has been previously investigated by the authors [16], who reported the successful incorporation of strontium into MBG framework and demonstrated, beside the excellent biocompatibility, the pro-osteogenic role of released $Sr^{2+}$ ions [16]. Moreover, with the aim to further expand the therapeutic potential, the co-substitution of different ions has been also explored [17], by producing Sr-Cu containing MBGs, where the pro-osteogenic effect of strontium was combined with the ability of copper to promote neovascularisation and to exert anti-microbial action [18,19].

In addition, the high number of terminal hydroxyl groups allows to easily functionalise MBG surface by grafting alkoxysilane moieties with the aim to achieve several targets, such as improved drug loading ability, the reduction of particle aggregation and nonspecific surface adhesion. In this regard, it is well known that the formation of nonspecific protein layer on the surface of devices contacted with biological fluids can heavily reduce their performance, due to the inhibitive effect on drug/ion release and on the formation of apatite layer [20], essential for a strong bond with living bone tissue and the promotion of osteogenesis. The adsorption of serum proteins has also a role on the microbial adhesion and biofilm formation [21,22], as the initial attachment of individual bacterial cells or small bacterial aggregates, is usually preceded by the adsorption of biological macromolecules [11,23], such as proteins. For this reason, the design of novel anti-fouling biomaterials able to prevent bone infections through the inhibition of protein adhesion and, consequently, bacterial attachment and colonisation would represent an attractive and promising strategy.

In this perspective, *zwitterionization* [5] has been extensively exploited and proposed as one of the most promising approaches to design anti-adhesive/anti-fouling surfaces [24,25] with hydrophilic behaviour and electrically neutral charges. *Zwitterionic* surfaces are characterized by an equal number of both positively and negatively charges in order to preserve the overall electrical neutrality [26] and the related anti-adhesive properties are imparted by a strongly bonded water molecule layer which acts as barrier against the adsorption of both proteins and bacteria [5,26]. In fact, the presence of a thick hydration layer allows proteins to preserve a stable conformation when approaching the substrate surface, avoiding irreversible adsorption [22,27,28].

In the contribution, in order to provide mixed anionic/cationic charges on the surface of the MBGs under physiological conditions (pH=7.4), maintaining a general electrical neutrality, a straightforward and clean methodology based on co-grafting of fully hydrolysable short-chain silane moieties is proposed as a reliable and effective alternative to previously reported approaches involving several synthetic steps and different intermediate products [29]. In particular, Sr-substituted MBGs (Sr-MBG) have been produced in the form of nano- and micro-particles, to obtain materials with different morphological and textural features, and subjected to a co-grafting reaction to introduce almost same amount of -$NH_3^+$/-$COO^-$ surface groups, expected to impart effective anti-adhesive properties. The functionalisation route based on the post-grafting of aminopropyl silanetriol (APST) and carboxyethylsilanetriol (CES), bearing respectively amino and carboxylate groups, was optimized based on the different reaction kinetics of the two precursors, both in terms of concentration and addition time. Fourier Transform Infrared Spectroscopy (FTIR) analyses allowed to evidence the successful grafting of -$NH_3^+$/-$COO^-$ groups and ζ-potential measurements permitted to establish the *zwitterionic* nature of the surface as function of pH.

In order to confirm that the post-functionalization does not inhibit the release property and bioactivity typical of unmodified Sr-MBGs, *in vitro* release of strontium ions was studied in Tris-HCl at pH 7.4 up to 14 days and *in vitro* bioactivity tests were performed in simulated body fluid (SBF) to investigate the deposition of apatite-like layer. *In vitro* biocompatibility assays in the presence of mouse pre-osteoblastic cell line MC3T3-E1 have been also performed to study the effect of post-grafted Sr-MBGs on the osteoblastic cell growth and differentiation. Finally, to assess the antifouling ability imparted by surface *zwitterionization*, reduced protein adsorption of serum proteins was evaluated by gel electrophoresis (SDS-PAGE) experiments using bovine serum albumin (BSA) and fibrinogen (Fib), respectively.

## 2. Materials and methods

### 2.1. Zwitterionic Sr-substituted MBGs

Sr-MBGs were prepared through two different synthesis routes to obtain nano- and micro-particles characterised by different textural features in term of exposed specific surface area, pore size and pore volume. The materials were post-functionalized by co-grafting aminopropyl silanetriol (APST) and carboxyethylsilanetriol (CES) as organosilane agents, whose relative amount and reaction time were properly adjusted to reach an almost neutral overall surface charge.

#### 2.1.1. Sr-substituted MBG nanoparticles

Sr-MBGs nanoparticles have been prepared by using a base-catalysed sol-gel synthesis, following the procedure reported previously by the authors[16]. In particular, MBGs with 2% molar percentage of Sr (molar ratio Sr/Ca/Si = 2/13/85, named hereafter as MBG_Sr2%_SG) was prepared as follows: 6.6 g cetyltrimethylammonium bromide (CTAB ≥98%, Sigma Aldrich, Italy) and 12 mL NH$_4$OH (Ammonium hydroxide solution, Sigma Aldrich, Italy) were dissolved in 600 mL of double distilled water (ddH$_2$O) under stirring for 30 min. Then, 30 mL tetraethyl orthosilicate (TEOS, Tetraethyl orthosilicate, Sigma Aldrich, Italy), 4.888 g of calcium nitrate tetrahydrate (Ca(NO$_3$)$_2$·4H$_2$O, 99%, Sigma Aldrich, Italy) and 0.428 g of strontium chloride (SrCl$_2$ 99%, Sigma Aldrich, Italy) were added under vigorous stirring for 3 h. The powder was collected by centrifugation (Hermle Labortechnik Z326) at 11,000 rpm for 5 min, washed one time with distilled water and two times with absolute ethanol. The final precipitate was dried at 70 °C for 12 h and calcined at 600 °C in air for 5 h at a heating rate of 1 °C min$^{-1}$ using a Carbolite 1300 CWF 15/5, in order to remove CTAB.

*2.1.2. Sr-substituted MBG microparticles*

Sr-MBGs in the form of microspheres with 2% molar percentage of Sr (molar ratio Sr/Ca/Si = 2/13/85, hereafter named as MBG_Sr2%_SD) were produce by using a spray-drying method, based on a modification of the procedure reported by Pontiroli et al. [30]. Briefly, 2.030 g of the non-ionic block copolymer Pluronic P123 (EO$_{20}$PO$_{70}$EO$_{20}$, average M$_n$ ~5,800, Sigma Aldrich, Italy) were dissolved in 85 g of ddH$_2$O. In a separate batch, a solution of 10.73 g of TEOS was pre-hydrolysed under acidic conditions using 5 g of an aqueous HCl solution at pH 2 until a transparent solution was obtained. The solution with TEOS was then added drop by drop into the template solution and kept stirring for 30 min. Then, 0.163 g of strontium chloride and 1.86 g of calcium nitrate tetrahydrate were added and the final solution was stirred for 15 min and then sprayed (Büchi, Mini Spray-Dryer B-290) using nitrogen as the atomizing gas with the following parameters: inlet temperature 220 °C, N$_2$ pressure 60 mmHg and feed rate 5 mL/min. The obtained powder was collected and calcined at 600 °C in air for 5 h at a heating rate of 1 °C min$^{-1}$ using a Carbolite 1300 CWF 15/5.

*2.1.3. Zwitterionic Sr-MBGs*

To provide *zwitterionic* properties, Sr-MBGs were post-grafted with 3-aminopropylsilanetriol (APST) (22-25% in water, ALFA Chemistry) and carboxyethylsilanetriol sodium salt (CES, 25% vol. in water, Carbosynth Limited). In particular, 0.5 g of dried powder were first outgassed overnight in vacuum and then dispersed in 100 mL of absolute ethanol. The initial amount of APTS and CES to be added to the suspension was calculated referring to the Zhuravlev number [31], which is widely used in the literature for the estimation of the organosilane precursors in post-grafting reactions of mesoporous silicas and MBGs. Indeed, according to the Zhuravlev number, a density 4.9 SiOH/nm$^2$ can be considered, and based on the measured specific surface area of MBGs, the overall number of exposed hydroxyls (acting as anchoring sites for grafting) was estimated and the amount of precursors derived accordingly. The optimisation of the overall procedure was not entirely straightforward and required several step-by-step adjustments. To this aim, ζ-potential measurements were conducted as preliminary evaluation of the modification of surface charge imparted by grafting, in order to guide the corrections required to reach the *zwitterionic* behaviour.

At first, an equimolar concentration of APST and CES (0.3 mmol) was added simultaneously to the ethanol suspension and refluxed 24 h under stirring (200 rpm) at 80 °C under nitrogen atmosphere. The resulting functionalized powders were filtered, washed three times with absolute ethanol and dried overnight at 70°C.

As reported in table 1 (first row), ζ-potential measurements at pH 7.4 revealed a negative value (-9 mV), suggesting higher surface reactivity for CES compared to APST. Therefore, to increase the amount of surface positive charges, a double amount of APST (0.6 mmol) was added to the ethanol solution and kept stirring for 30 min prior the addition of CES (0.3 mmol), in order to allow the reaction between MBG surface and APST. Lastly, an amount of APST equal to 0.9 mmol (molar ratio APST:CES 3:1) resulted to be the most appropriate to reach an overall surface charge close to zero, as shown by table 1 (third row). The samples reacted with APST:CES 3:1 molar ratio will be referred hereafter as MBG_Sr2%_SG_Z and MBG_Sr2%_SD_Z and were selected for the physico-chemical characterization and for the biological assessment. Fig. 1 displays functionalization procedure to obtain the *zwitterionic* materials.

Table 1 Investigated APST/CES amounts (mmol) and addition time. Zeta potential values measured at pH 7.4 of the resulting functionalized MBG_Sr2%.

| mmol APST | mmol CES | Addition time | ζ-potential pH 7.4 |
|---|---|---|---|
| 0.3 | 0.3 | Simultaneously | -9 mV |
| 0.6 | 0.3 | CES added after 30 min | -7 mV |
| 0.9 | 0.3 | CES added after 30 min | -2 mV |

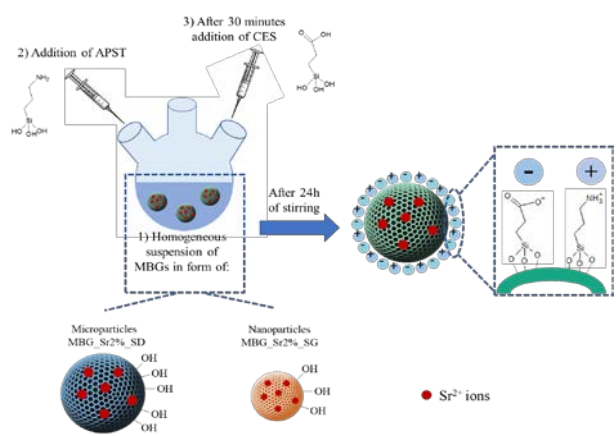

Fig. 1 Functionalization procedure to obtain *zwitterionic* Sr-MBGs.

*2.2. Physico-chemical characterization*

The mesoporous structure of materials was characterised by low angle X-ray diffraction (XRD) and transmission electron microscopy (TEM). XRD experiments were performed on a Philips X'Pert diffractometer equipped with a Cu Kα (40 kV, 20 mA) source over the range 0.6–8.0 θ, with a step of 0.02 θ and a contact time of 5 s. TEM images were recorded in a JEOL 3010 electron microscope (JEOL, Japan) operating at 300 keV coefficient of objective lens (Cs) equal to 0.6 mm, resolution 1.7 Å, employing a CCD camera (1024 x 1024 pixels, size 24 x 24 lm; MultiScan model 794, Gatan, UK) under low dose conditions.

The morphology of the produced particles was analysed by Field-Emission Scanning Electron Microscopy (FE-SEM) using a ZEISS MERLIN instrument. For FE-SEM observations, 10 mg of MBG_Sr2%_SG_Z were dispersed in 10 mL of isopropanol using an ultrasonic bath (Digitec DT 103H, Bandelin) for 5 min to obtain a stable suspension. The resulting suspension was dropped on a copper grid (3.05 mm Diam.200 MESH, TAAB), allowed to dry and successively chromium-coated prior to imaging (Cr layer of *ca* 7 nm). At variance, MBG_Sr2%_SD_Z powder was dispersed directly to a conductive carbon tape adhered on a stub and coated with the Cr layer (7 nm).

Textural properties were analysed by $N_2$ adsorption-desorption measurement conducted by ASAP2020, Micromeritics analyser at a temperature of –196 °C and before measurements, samples were outgassed at 150 °C for 5 h. The Brunauer-Emmett-Teller (BET) equation was used to calculate the specific surface area ($SSA_{BET}$) from the adsorption isotherm in the 0.04–0.2 relative pressure range. The mesoporous silica pore size distribution was calculated through the DFT method (Density Functional Theory) using the NLDFT kernel of equilibrium isotherms (desorption branch).

The successful anchoring of functional groups was assessed by Fourier Transform infrared spectroscopy (FT-IR), thermo-gravimetric analysis (TGA) and ζ potential measurements. FT-IR spectra were collected using a FT-IR spectrometer (Bruker Equinox 55 spectrometer) in the 4000-400 $cm^{-1}$ wavenumber range. TGA were conducted on a TG 209 F1 Libra instrument (Netzsch) over a temperature range of 25–600 °C with a heating rate of 10 °C/min under air in a flow of 50 mL/min).

ζ-potential measurements (Zetasizer nano ZS90 Malvern Instruments Ltd.) were conducted in aqueous media at different pH values in order to determine in which pH range the material surface preserves *zwitterionic* nature, i.e. the isoelectric point, which is closely related to the zero point charge [32].

*2.3. In vitro bioactivity of zwitterionic MBGs*

*In vitro* bioactivity test was performed in simulated body fluid (SBF), according to the protocol described in the literature [33], with the aim to evaluate the apatite-forming ability of the functionalized Sr-substituted MBGs. To this aim, 30 mg of Sr-MBGs were soaked in 30 mL of SBF at 37 °C up to 14 days in an orbital shaker (Excella E24, Eppendorf, Milan, Italy) with an agitation rate of 150 rpm. At each time point (3 h, 1 day, 3 days, 7 days and 14 days), the suspension was centrifuged at 10,000 rpm for 5 min, the collected powder was washed twice with distilled water and dried in oven at 70 °C for 12 h prior FE-SEM and XRD analysis to evaluate the apatite layer formation. Moreover, the pH of each recovered supernatant was measured, to assess if the values are suitable for allowing osteoblasts to maintain their physiological activity [34].

*2.4. $Sr^{2+}$ release from zwitterionic MBGs*

The concentration of $Sr^{2+}$ ions released from *zwitterionic* MBGs was evaluated by soaking the powders in Tris HCl buffer (Tris(hydroxymethyl)aminomethane (Trizma) (Sigma Aldrich, Milan, Italy) 0.1 M, pH 7.4) at concentration of 250 μg/mL[16,19]. In particular, 5 mg of powder were suspended in 20 mL of buffer up to 14 days at 37 °C in an orbital shaker (Excella E24, Eppendorf) with an agitation rate of 150 rpm. At defined time points (3 h, 24 h, 3 days, 7 days and 14 days) the suspension was centrifuged at 10,000 rpm for 5 min (Hermle Labortechnik Z326, Wehingen, Germany), half of the supernatant was collected and replaced by the same volume of fresh buffer solution to keep constant the volume of the release medium. The release experiments were carried out in triplicate. The concentration of $Sr^{2+}$ ions was measured by Inductively Coupled Plasma Atomic Emission Spectrometry Technique (ICP-AES) (ICP-MS, Thermoscientific, Waltham, MA, USA, ICAP Q), after appropriate dilutions. To evaluate the effective amount of strontium incorporated into MBGs during the synthesis, the powders were dissolved in a mixture of nitric and hydrofluoric acids (0.5 mL of $HNO_3$ and 2 mL of HF for 10 mg of powder) and the resulting solutions were measured via ICP analysis.

*2.5. Biological assessment*

*2.5.1. In vitro biocompatibility tests*

Cell culture experiments were performed using the mouse MC3T3-E1 osteoblast cell line, plated at an initial density of 10,000 cells cm$^{-2}$ in a multi-well plates and incubated in 2 mL of Dulbecco's modified Eagle's medium (DMEM) (Sigma Chemical Co., St. Louis, USA) supplemented with 10% heat-inactivated fetal bovine serum (FBS) (Lonza BioWhittaker, Spain), 1 mM L-glutamine (Lonza BioWhittaker, Spain), 200 µg ml$^{-1}$ penicillin (Lonza BioWhittaker, Spain) and 200 µg ml$^{-1}$ streptomycin (Lonza BioWhittaker, Spain) at 37°C in 5% $CO_2$.

Sr-substituted MBGs (before and after functionalisation) were suspended at different concentrations (25, 50 and 75 ug/mL) in completed DMEM, sonicated in order to obtain a stable suspension and placed into each 24-well plates after cell seeding. Wells without MBG particles were used as control.

*2.5.2. Mitochondrial activity – MTT*

Cell proliferation induced by both unmodified and *zwitterionic* Sr-substituted MBGs was determined by MTT test. MC3T3-E1 cells were seeded for 24 h as previously described and then incubated at 37°C in 5% $CO_2$ with different particle concentrations (25, 50 and 75 ug/mL) for different time periods (1 day, 2 days and 5 days). At each time point, 300 µL of CellTiter 96® AQueous One Solution Reagent (containing 3-(4,5-dimethythizol-2-yl)-5-(3-carboxymethoxyphenyl)-2-(4-sulfophenyl)-2H-tetrazolium salt (MTS) and an electron coupling reagent (phenazine ethosulfate) that combines with MTS to form a stable solution) was added to each well and incubated for 2 h. then, 100 µl of each cell conditioned medium was placed in a 96-well plate and the absorbance at 490 nm was measured in an Opsys MR Reader (Dynex Technologies, Chantilly, VA, USA).

*2.5.3. Citotoxicity test –Lactate dehydrogenase (LDH) measurement*

Biocompatibility and cytotoxic effects of unmodified and *zwitterionic* Sr-substituted MBGs were tested measuring the amount of released cytosolic lactate dehydrogenase (LDH) enzyme caused by the cellular membrane damage (Spinreact S.A., Spain). The assay is based on the reduction of NAD by LDH. The resulting reduced NAD (NADH) is exploited in the stoichiometric conversion of a tetrazolium dye.

The supernatants of particles-cells incubation experiments were collected after 1 day of incubation and mixed with the LDH kit reagents (SPINREACT®, imidazole 65 mmol/L, pyruvate 0.6 mmol/L and NADH 0.18 mmol/L). The absorbance of the resulting coloured compound was measured at 340 nm wavelength at 0, 1, 2, and 3 min, calculating the amount of enzyme that transforms 1 mmol of substrate per minute in standard conditions (U/L) as ΔA/min x 4925 (where ΔA is the average difference per minute and 4925 corresponds to the correction factor for 25–30 °C).

*2.5.4. Mineralization assay*

Matrix mineralization was measured in MC3T3-E1 cell cultures by alizarin red staining, as described in the protocol reported by Lozano et al. [35]. After incubation with the tested materials for 15 days, cells were washed three times with PBS and then fixed with 75% ethanol for 1 h. Cell cultures were stained with 40 mM alizarin red in distilled water (pH 4.2) for 10 min at room temperature. Subsequently, cell monolayers were washed five times with distilled water and the stain was dissolved with 10% cetylpyridinum chloride in 10 mM sodium phosphate, pH 7, measuring absorbance at 620 nm.

*2.6. Anti-fouling abilities assessment: serum protein adhesion assay*

*In vitro* adhesion of proteins was determined using bovine serum albumin (96% BSA, A2153 Sigma-Aldrich) and fibrinogen (75% Fib, F8630 Sigma-Aldrich). A 50% v/v solution of such proteins (2 mg/mL for BSA and Fib in PBS 1x, respectively) was gently mixed with both unmodified and *zwitterionic* Sr-MBGs (2 mg/mL in PBS 1x) and kept under orbital agitation (200 rpm) during 24h at 37 ºC. Particle suspensions were then centrifuged (10,000 rpm) and washed with fresh PBS to eliminate free or loosely bound proteins and a one-dimensional sodium dodecyl sulfate-polyacrylamide gel electrophoresis gel (SDS-PAGE) assay was performed. Briefly, the recovered supernatants were mixed with a buffer (Tris 6 mM, SDS 2%, glycerol 10%, 2-β-mercaptoethanol 0.5 M, traces of bromophenol blue, pH = 6.8) and loaded in 10% SDS-PAGE gels. A calibration curve was estimated by loading 0.5, 1.0, 1.5, 2.0 and 2.5 µg/mL protein concentrations. The gels were run with a constant 100 V voltage (45-60 min) and stained in R-250 colloidal Coomasie blue solution for visualization. Imagen J software was used in order to determine the amount of adsorbed proteins through calibrated line of the different proteins. Three different measurements of two independent experiment were performed in order to determine the statistical analyses.

## 3. Results and discussion

*3.1. Physico-chemical characterization*

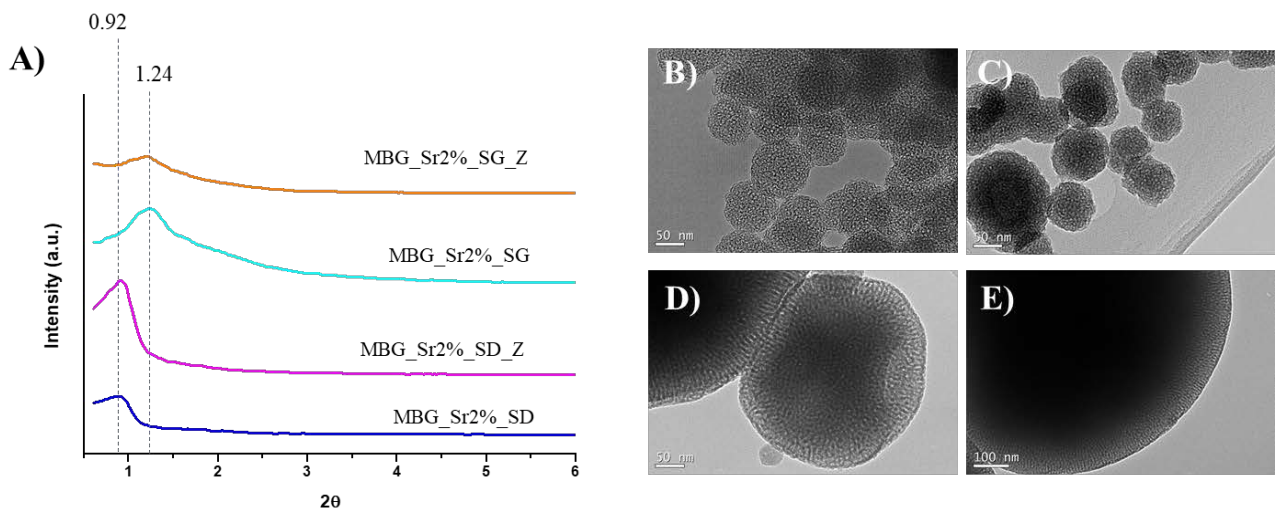

Fig. 2 (A) Low-angle XRD of unmodified and zwitterionic Sr-MBGs. TEM images of: (B) MBG_Sr2%_SG; (C) MBG_Sr2%_SG_Z; (D) MBG_Sr2%_SD; (E) MBG_Sr2%_SD_Z.

The mesostructure of unmodified and *zwitterionic* Sr-MBGs has been evaluated by TEM and low-angle XRD (Fig 2). TEM images evidenced a worm-like mesoporosity throughout both nanoparticles (Fig 2.B) and microparticles, respectively (Fig 2.D), also clearly discernible after the functionalization (Fig 2.C-E). Low-angle XRD patterns (Fig 2.A) confirm the worm-like porous structure [36–38], in fact both MBG_Sr2%_SG and MBG_Sr2%_SD, before and after functionalization, showed a single peak, respectively at around 1.24° and 0.92°, corresponding to the (100) reflection. The decrease of the reflection intensity revealed upon silane co-grafting can be ascribed to a limited and confined collapse of the mesoporous structure.

Morphological analysis of MBG_Sr2%_SG_Z showed particles with uniform spherical shape and size ranging between 100 and 200 nm, while MBG_Sr2%_SD_Z, revealed spherical particles with micrometric size between 0.5 and 5 µm (Fig SI 1). These results resulted in good agreement with morphology, size and shape revealed by the observation of unmodified analogues materials[16], evidencing that the post-synthesis modification did not significantly alter the morphological features of Sr-MBGs.

ICP-AES analysis conducted on acid-digested powder revealed that only 30% of strontium precursor was effectively incorporated for MBG_Sr2%_SG, at variance with MBG_Sr2%_SD samples whose incorporated strontium amount resulted very similar to the theoretical value. ICP analysis on samples after the post-synthesis reaction evidenced that incorporated strontium amount resulted unaffected, demonstrating that the functionalization reaction did not induce any loss of incorporated strontium amount.

The textural features of the samples before and after functionalization were investigated by $N_2$ adsorption/desorption measurements. MBG_Sr2%_SG showed a type IV isotherm, characteristic of mesoporous materials and a pore size distribution centred at around 4.2 nm (Fig 3.A-B).

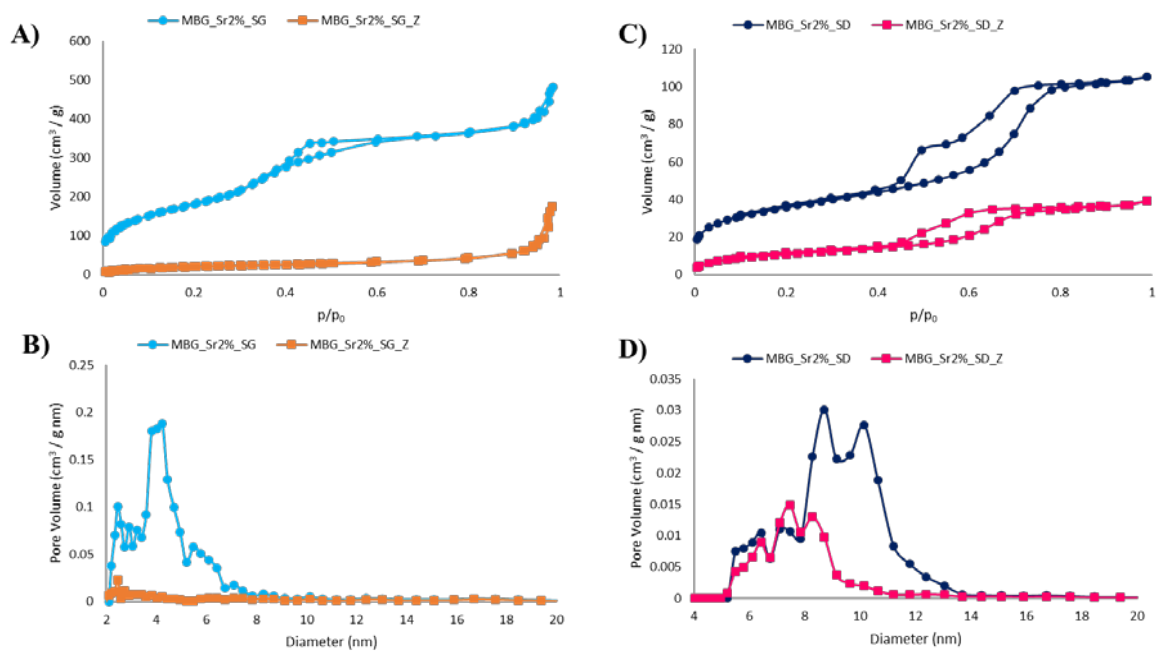

Fig. 3 $N_2$ adsorption-desorption isotherm of MBG_Sr2%_SG and MBG_Sr2%_SG_Z (A), MBG_Sr2%_SD and MBG_Sr2%_SD_Z (C). DFT pore size distribution of MBG_Sr2%_SG and MBG_Sr2%_SG_Z (B), MBG_Sr2%_SD and MBG_Sr2%_SD_Z (D).

The material exhibits excellent textural properties, in terms of very high surface area (670 m² g$^{-1}$) and pore volume (0.63 cm³ g$^{-1}$), as reported in Table 2. After the functionalization, a modification of the isotherm and a drastic decrease in SSA and pore volume (Table 2) is observed, suggesting that mesopore entrances are partially or even fully blocked by the organosilane grafting [38,39]. In fact, if the organosilanes react at the pore entrances during the initial phases of reaction, the diffusion of precursors within the mesopore can be compromised, leading to an uneven distribution and, in some cases, to the almost closure of the pores [39].

$N_2$ adsorption-desorption isotherm of MBG_Sr2%_SD is a IV type isotherm with a pronounced hysteresis loop. The related values of specific surface area SSA, pore volume and pore size are reported in table 2 and are in accordance to those previously obtained for analogue systems [16]. As expected after grafting, MBG_Sr2%_SD_Z exhibits lower specific surface area and pore volume and size compared with unmodified sample, but at variance with MBG_Sr2%_SG_Z, a noticeable residual surface area and pore volume is retained after silane anchoring, evidencing that pores do not undergo complete occlusion but experienced a size reduction (from 8-11 nm to 6-9 nm).

Table 2 Textural parameters of MBG_Sr2%_SG, MBG_Sr2%_SG_Z, MBG_Sr2%_SD and MBG_Sr2%_SD_Z.

|  | MBG_Sr2%_SG | MBG_Sr2%_SG_Z | MBG_Sr2%_SD | MBG_Sr2%_SD_Z |
|---|---|---|---|---|
| **BET surface area** | 670 m² g$^{-1}$ | 76 m² g$^{-1}$ | 126 m² g$^{-1}$ | 40 m² g$^{-1}$ |
| **Average Pore size** | 4.2 nm | - | 8-11 nm | 6-9 nm |
| **Pore volume** | 0.63 cm³ g$^{-1}$ | 0.16 cm³ g$^{-1}$ | 0.16 cm³ g$^{-1}$ | 0.06 cm³ g$^{-1}$ |

FT-IR spectroscopy allowed to assess the successful anchoring of functional groups. Figure 4 shows the FT-IR spectra of both bare and *zwitterionic* samples. All spectra display in the range of 3750–3000 cm$^{-1}$ the typical absorption bands, corresponding to the stretching vibrational frequencies of the H-bonded hydroxyls. The spectra of functionalized samples show significant changes compared to the bare analogues, in particular, two bands at 1550 and 1407 cm$^{-1}$, ascribed respectively to the asymmetric ($\nu_{as}$) and symmetric ($\nu_s$) stretching vibration of the carboxylate group COO$^-$ [40], and two shoulders at around 1650 cm$^{-1}$ and 1520 cm$^{-1}$ corresponding to the bending mode of protonated amine group NH$_3^+$ [41]. On contrary, the C=O adsorption band of protonated COOH group, displayed at 1706 cm$^{-1}$, and the band at 1595 cm$^{-1}$, ascribable to the bending mode of neutral -NH$_2$ do not appear in the FT-IR spectrum of *zwitterionic* samples. Taken together these FT-IR observations reveal that MBG_Sr2%_SG_Z and MBG_Sr2%_SD_Z exhibits a mixed charged surface, due to the co-presence of NH$_3^+$ and COO$^-$ groups, respectively.

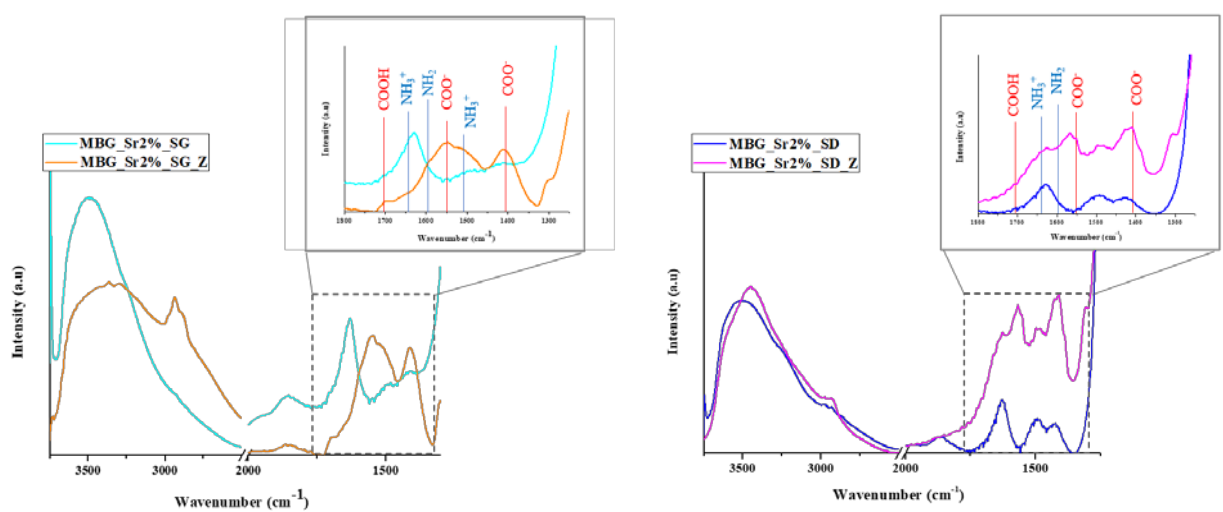

Fig. 4 FTIR spectra of MBG_Sr2%_SG and MBG_Sr2%_SD before and after *zwitterionization*.

Thermogravimetric analysis has been used to estimate the anchored organic components based on the weight loss. As reported in Fig SI 2, TG profiles of bare Sr-MBGs show solely a weight loss in the range 30-180°C, related to the elimination of the adsorbed water, revealing the absence of additional surface moieties. At variance, *zwitterionic* samples, two clear weight loss components were observed: the first below 200° C due to the release of the adsorbed water, and a more significant step between 300 and 600°C, ascribable to the decomposition of anchored amino and carboxylate organic moieties.

Since these materials are intended for biomedical applications where the surface charge plays an essential role, their behaviour in aqueous media at different pH was investigated by ζ-potential measurements, in order to determine the pH values at which the *zwitterionic* nature is preserved. As can be observed from the ζ-potential *vs.* pH plots reported in Fig 5, unmodified Sr-MBGs exhibited negative ζ-potential values within the whole pH range, due to the presence of deprotonated silanols (-Si –O$^-$). These data are in good agreement with those reported in the literature, where the isoelectric point at around pH = 2 is reported for similar silica-based systems [42,43].

As already mentioned before, ζ-potential analysis supported the optimisation of the functionalization protocol in term of organosilane initial ratio and reaction times. The first attempt conducted with an equimolar concentration of APST and CES, added simultaneously (MBG_Sr2%_Z_1:1) led to the ζ-potential value at pH 7.4 around -9 mV. This negative value can be ascribed both to a preferential grafting of CES with respect to APST or the presence of residual unreacted deprotonated silanols groups, leading to a larger population of negative charged (COO$^-$ and SiO$^-$), compared to protonated amino groups (-NH$_3^+$).

A double concentration of APTS, followed by CES addition after 30 minutes of reaction (MBG_Sr2%_Z_1:2), allowed to obtain lees negative ζ-potential value at pH 7.4 (-6 mV), thanks to a larger amount of amino groups grafted on the particle surface. Finally, a further increase of APTS concentration on equal addition time of CES led to a closer value of ζ-potential to a global zero charge, given by the balance of positive and negative charges. ζ-potential measurements of MBG_Sr2%_SG_Z and MBG_Sr2%_SD_Z (MBG_Sr2%_Z_1:3) reveals positive values in the pH range 3-7, due to the protonation of amino groups (NH$_2$ > NH$_3^+$), which reach a value close to zero (2.2 ± 0.9 mV and -2.8 ± 1.2 mV, respectively) at pH 7.4, which indicates the *zwitterionic* behaviour of both systems in physiological conditions.

Since MBG_Sr2%_SG_Z and MBG_Sr2%_SD_Z samples exhibited the required $NH_3^+$/-$COO^-$ *zwitterionic* pairs and net surface charges close to zero at the physiological pH of 7.4, the biocompatibility and the capability to inhibit protein adhesion of these two functionalized samples have been tested and compared to the bare analogues.

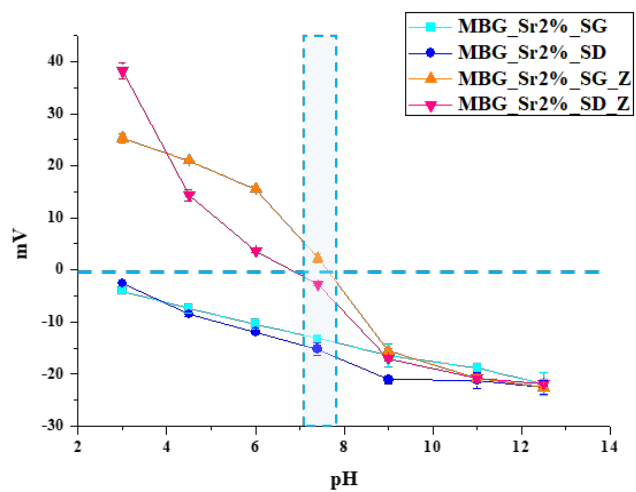

Fig. 5 ζ-potential measurements of MBG_Sr2%_SG and MBG_Sr2%_SD recorded at different pH before and after *zwitterionization*.

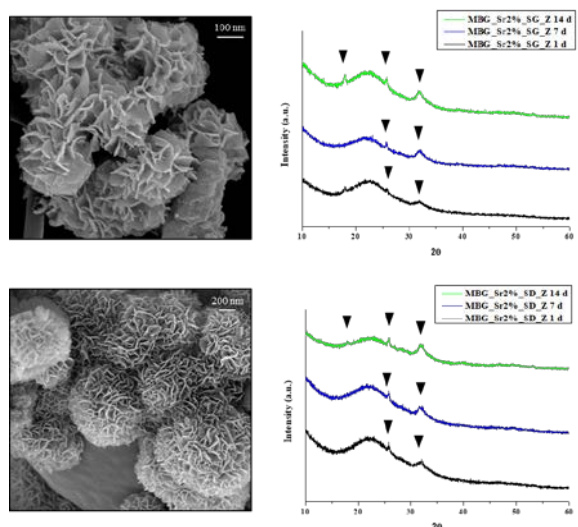

Fig. 6 FESEM images and XRD patterns of MBG_Sr2%_SG_Z and MBG_Sr2%_SD_Z after 1d, 7d and 14d of soaking in SBF (the most representative diffraction peaks of apatite phase are highlighted). The FESEM images are related to the surface after 14 days of incubation.

### 3.2. *In vitro bioactivity of zwitterionic Sr-MBGs*

After the grafting of APST and CES, both Sr-MBGs show a remarkable bioactive behaviour, retaining their ability to induce the deposition of the HA layer with fast kinetics [44]. In fact, after only 1 day of soaking, irrespective of the synthesis route, both Sr-MBGs appeared covered by a rough layer of globular agglomerates of apatite-like phase, that grew in size during the test, causing the final embedding of MBG particles overtime. As shown in Fig. 6, both particles after 7 days were fully covered by a compact layer of needle-like nanocrystals with the characteristic cauliflower morphology. EDS analysis performed on dried powders evidenced a Ca/P ratio of 1.7, typical value reported in the literature for carbonated hydroxyapatite [45,46]. Furthermore, XRD patterns confirmed the formation of apatite-like layer with nanocrystalline nature (Fig 6), as revealed by peaks at 25.8 and 32.0 2θ, matching the HA reference (00-001-1008).

Finally, at each time point, the pH of the SBF solution was measured, resulting along the overall duration below 7.8, which allows osteoblasts to maintain their physiological activity[47].

The preservation of Sr-MBG bioactivity after *zwitterionization* was not fully predictable as the ion-exchange reactions which promote HA deposition could have been hampered or even fully blocked by the presence of the silane moieties. The assessment of this aspect was one the major goal of the work as it represents an essential feature for promoting bone regeneration.

### 3.3. $Sr^{2+}$ release from zwitterionic MBGs

The ion release of bare and functionalized samples was evaluated in Tris-HCl medium (pH 7.4); samples were incubated at 37 °C up to 14 days and, at selected time points (3 h, 1 day, 3 days, 7 days and 14 days), were centrifuged, aliquots were withdrawn and analysed by ICP-AES. As expected, the release properties of Sr-MBGs before functionalization resulted fully in agreement with those already reported by the authors in a previous contribution, confirming the fast ion diffusion inside the porous structure [16,19].

Strontium release from *zwitterionic* Sr-MBGs, evaluated in the same experimental conditions (Fig SI 3) revealed that samples after functionalization were able to release the total amount of incorporated $Sr^{2+}$ ions with kinetics to those shown by the corresponding bare samples [16]. The final release $Sr^{2+}$ concentration (1.6 ppm for MBG_Sr2%_SG_Z and 4.4 ppm for MBG_Sr2%_SD_Z) has the potential to stimulated osteogenic response as shown in several work, without inducing any cytotoxic effect [48]. In this regard [16], the authors recently demonstrated that the same concentration of $Sr^{2+}$ was able to stimulate the expression of pro-osteogenic genes (COLL1A1, SPARC and OPG) of osteoblast-like cells cultured in the presence of only the dissolved ions released from Sr-substituted particles, confirming the key role exerted by strontium in stimulating osteoblast cell activity. Newly, the dissolution products released by Sr-containing bioactive glass nanoparticle were successfully proved to induce the stimulation of osteogenic response in hMSCs with the expression of genes associated to early-, mid- and late-osteogenic marker in the absence of osteogenic supplements [48].

*1.4 Cell viability*

The use of MBG nanocarriers for clinical applications requires excellent biocompatibility and absence of any cytotoxicity. In the literature, silica-based bioactive glass particles were widely proved to be biocompatible materials that exhibits low toxicity and lack of immunogenicity, degrading into nontoxic compounds (mainly silicic acid) in relatively short time periods[16,49–51].

Despite this lack of toxicity, the surface modification could provoke the appearance of toxicity due to several consequent aspects, among the other an enhanced uptake within the cells or the release of functional groups degradation products. To evaluate cytotoxicity, MC3T3-E1 preosteoblast cells were incubated with different amount of the both SD and SG before and after zwitterionization in cell culture medium. At determined incubation times (1, 2 and 5 days), cell viability was evaluated via the standard cell viability test by MTS reduction. The results showed that the tested particles, irrespective of their size and surface features, did not exhibit any cytotoxicity in preosteoblastic MC3T3-E1 (Fig 7) [52]. In order to further confirm these results lactate dehydrogenase (LDH) production were recorded after 1 day of cell incubation with the different Sr-MBG particles (before and after *zwitterionization*) at different concentrations. LDH is an enzyme released by cells in case of cell membrane rupture, thus indicating a cytotoxic effect. Interestingly, the results of LDH tests reported (Fig SI 4 A) obtained for MBG_Sr2%_SG and MBG_Sr2%_SD and the *zwitterionic* analogues, evidenced no cytotoxic effect up to the concentration of 75 µg/mL and no significant differences were registered in comparison with the control test. These results were also confirmed by optical microscopy (Fig SI 4 B), where the cells adhered to the surface of the well appeared well spread without any apparent cell damage.

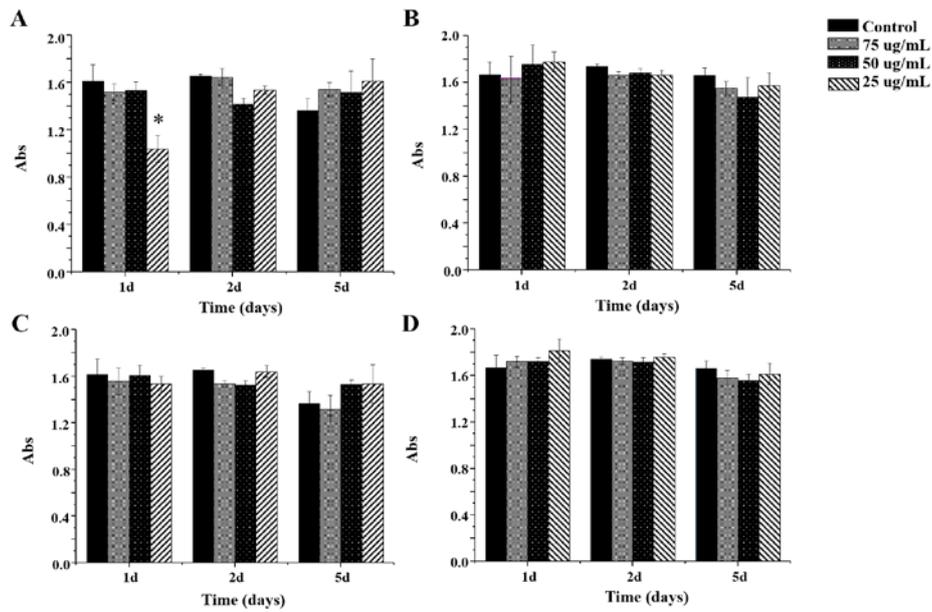

Fig. 7 Cell viability studies of the samples MBG_Sr2%_SG (A), MBG_Sr2%_SG_Z (B), MBG_Sr2%_SD (C) and MBG_Sr2%_SD_Z (D) at different concentration for MC3T3-E1 cell line and 1 d, 2 d and 5 d of exposure time. *p <0.05 *vs* corresponding control without particles (ANOVA)

*3.5 Mineralization assay*

Differentiation is a process by which unspecialized cells, pre-osteoblasts, become specialized cells with the function to restore the bone. In this study a preliminary study by monitoring the mineralization process in term of alizarin assay has been carried out to clarify if the zwitterionization process affect or not to the differentiation. Related results, reported in Fig 8 evidenced an increase of matrix mineralization for all Sr-substituted MBGs materials with respect to control, which is attributed to the osteogenic capability of released $Sr^{2+}$ ions release as it has been previously reported [16, 48, 53]. After functionalization, no significant differences with respect to both pristine samples were observed, confirming that the zwitterionization process does not affect to alizarin production. These findings could be explained since the functionalized samples exhibit similar strontium release than pristine samples, which does not affect to the associated osteogenic capability (in this case the mineralization).

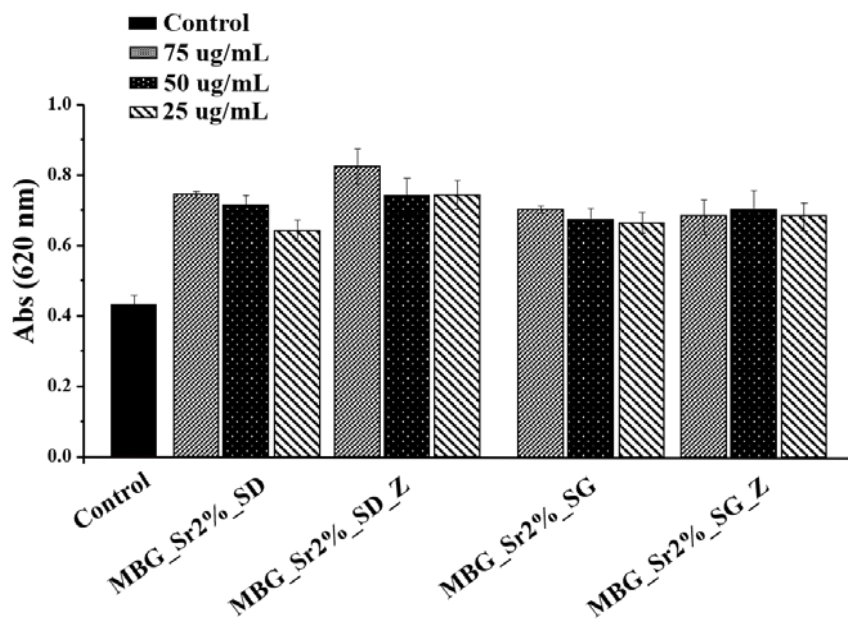

Fig. 8 Mineralization process in terms of alizarin assays of Sr-MBGs (SD and SG) before and after *zwitterionization* at different concentration (25, 50 and 75 μg/mL).

*3.6 In vitro non-specific protein adhesion*

Once demonstrated the *zwitterionic* nature and the biocompatibility of the functionalised Sr-MBG particles, their tendency to undergo nonspecific protein adsorption when soaked in biological fluids was evaluated by *in vitro* adsorption assays with two serum proteins, BSA and fibrinogen, respectively. Related results, displayed in Table 3, show a notable decrease of protein adsorption for both proteins after *zwitterionization* process with a reduction of *c.a.* 85% and 86% of BSA and 70% and 54% of Fib for the MBG_Sr2%_SG_Z and MBG_Sr2%_SD_Z, respectively.

Table 3 Amount of BSA and fibrinogen adsorbed on the surfaces of Sr-MBGs before and after *zwitterionization* determined by SDS-PAGE technique.

| Sample | BSA (mg/g) | Fibrinogen (mg/g) |
|---|---|---|
| **MBG_Sr2%_SG** | 348 ± 75 | 463 ± 79 |
| **MBG_Sr2%_SG_Z** | 51 ± 18 | 134± 58 |
| **MBG_Sr2%_SD** | 170± 50 | 243± 58 |
| **MBG_Sr2%_SD_Z** | 24± 13 | 133± 57 |

This significant reduction is attributed to the presence of *zwitterionic* pairs, which are known to act as a barrier to non-specific protein adsorption via a strongly adsorbed layer of water molecules solvating the charged terminal $COO^-$/$NH_3^+$ groups via multiples hydrogen bounds.

The obtained data reveal a significant larger reduction in protein adhesion for MBG_Sr2%_SG_Z compared MBG_Sr2%_SD_Z, when both compared to their corresponding bare samples, that can be attributed to the drastic decrease of surface area upon functionalization evidenced by nitrogen adsorption-desorption analysis (Table 2) for Sr-MBG prepared by SG approach.

## 4. Conclusions

In this study, promising advanced biomaterials based on strontium-releasing mesoporous bioactive glasses with low-fouling *zwitterionic* surface were developed to promote bone regeneration and simultaneously inhibit non-specific biomolecules adsorption and, consequently, bacterial adhesion. To reach this goal, Sr-MBGs have been produced in form of nano- and micro-particles and modified by co-grafting of hydrolysable short chain silane bearing amino (aminopropyl silanetriol, APST) and carboxylic (carboxyethylsilanetriol, CES), respectively. A straightforward and clean procure has been optimized by modulating the relative amount of APST and CES and their reaction time, to provide Sr-MBG surface with an overall electrical neutrality under physiological conditions (pH=7.4).

The modification of the surface with *zwitterionic* pairs was proved to not alter the ability to release strontium ions within concentrations able to promote early osteogenic differentiation and mineralized matrix deposition, without any cytotoxic effect as demonstrated by cell culture with MC3T3-E1 preosteoblast cells.

In *vitro* non-specific protein adhesion assay demonstrated that the successful grafting of the *zwitterionic* pairs onto Sr-MBG surfaces reduces BSA and Fib adhesion up to *ca.* 85% and 70%, respectively.

## 5. Acknowledgements

This work was supported by Compagnia di San Paolo under the initiative "Metti in rete la tua idea di ricerca". M.V.R. and I.I.B. acknowledge funding from the European Research Council (Advanced Grant VERDI; ERC-2015-AdG Proposal No. 694160).

## 6. References

[1]    T. Winkler, F.A. Sass, G.N. Duda, K. Schmidt-bleek, A review of biomaterials in bone defect healing , remaining shortcomings and future opportunities for bone tissue engineering, Bone Joint Res 7 (2018) 232-243.

[2]    N.P. Haas, Callusmodulation Fiktion oder Realität?, Chirurg 71 (2000) 987-988.

[3] C. Schlundt, C.H. Bucher, S. Tsitsilonis, H. Schell, G.N. Duda, K. Schmidt-bleek, Clinical and Research Approaches to Treat Non-union Fracture, Current Osteoporosis Reports (2018) 155-168.

[4] D. Campoccia, L. Montanaro, C. Renata, Biomaterials A review of the biomaterials technologies for infection-resistant surfaces, Biomaterials. 34 (2013) 8533-8554.

[5] I. Izquierdo-Barba, M. Colilla, M. Vallet-Regí, Zwitterionic ceramics for biomedical applications, Acta Biomater. 40 (2016) 201-211.

[6] S. Sànchez-Salcedo, M. Colilla, I. Izquierdo-Barba, M. Vallet-Regí, Design and preparation of biocompatible zwitterionic hydroxyapatite, J. Mater. Chem. B. (2013) 1595-1606.

[7] J. Mader, M.E. Shirtliff, S.C. Bergquist, J. Calhoun, Antimicrobial Treatment of Chronic Osteomyelitis, Clin. Orthop. Relat. Res. (1999) 47-65.

[8] R.O. Darouiche, Treatment of Infections Associated with Surgical Implants, N. Engl. J. Med. 350 (2004) 1422-1429.

[9] R. Haidar, A. Der Boghossian, B. Atiyeh, Duration of post-surgical antibiotics in chronic osteomyelitis : empiric or evidence-based?, Int. J. Infect. Dis. 14 (2020) e752-e758.

[10] J.D. Whitehouse, N.D. Friedman, K.B. Kirkland, W.J. Richardson, D.J. Sexton, The impact of surgical-site infections following orthopedic surgery at a community hospital and a university hospital: adverse quality of life, excess length of stay, and extra cost, Infect. Control Hosp. Epidemiol. 23 (2012) 183-189.

[11] J.W. Costerton, P.S. Stewart, E.P. Greenberg, Bacterial Biofilms : A Common Cause of Persistent Infections, Science 284 (1999) 1318-1323.

[12] R. García-Alvarez, I. Izquierdo-Barba, M. Vallet-Regí, 3D scaffold with effective multidrug sequential release against bacteria biofilm, Acta Biomater. 49 (2017) 113-126.

[13] R.M. Donlan, J.W. Costerton, Biofilms : Survival Mechanisms of Clinically Relevant Microorganisms, Clin. Microbiol. Rev. 15 (2002) 167-193.

[14] J.W. Costerton, L. Montanaro, C.R. Arciola, Biofilm in Implant Infections: Its Production and Regulation, Int. J. Artif. Organs. 28 (2005) 1062-1068.

[15] D. Campoccia, F. Testoni, S. Ravaioli, I. Cangini, A. Maso, P. Speziale, L. Montanaro, L. Visai, C.R. Arciola, Orthopedic implant infections : Incompetence of Staphylococcus epidermidis , Staphylococcus lugdunensis , and Enterococcus faecalis to invade osteoblasts, J. Biomed. Mater. Res 104 (2015) 788-801.

[16] S. Fiorilli, G. Molino, C. Pontremoli, G. Iviglia, E. Torre, C. Cassinelli, M. Morra, C. Vitale-Brovarone, The incorporation of strontium to improve bone-regeneration ability of mesoporous bioactive glasses, Materials 11 (2018) 678-696.

[17] A. Bari, G. Molino, S. Fiorilli, C. Vitale-Brovarone, Novel multifunctional strontium-copper co-substituted mesoporous bioactive particles, Mater. Lett. 223 (2018) 37-40.

[18] M. Shi, Z. Chen, S. Farnaghi, T. Friis, X. Mao, Y. Xiao, C. Wu, Copper-doped mesoporous silica nanospheres, a promising immunomodulatory agent for inducing osteogenesis, Acta Biomater. 30 (2016) 334-344.

[19] C. Pontremoli, M. Boffito, S. Fiorilli, R. Laurano, A. Torchio, A. Bari, C. Tonda-Turo, G. Ciardelli, C. Vitale-Brovarone, Hybrid injectable platforms for the in situ delivery of therapeutic ions from mesoporous glasses, Chem. Eng. J. 340 (2018) 103-113.

[20] S. Lin, J.R. Jones, The effect of serum proteins on apatite growth for 45S5 Bioglass and common sol-gel derived glass in SBF, Biomed. Glasses 4 (2018) 13-20.

[21] C. Blaszykowski, S. Sheikh, M. Thompson, Surface chemistry to minimize fouling from blood-based fluids, Chem. Soc. Rev. 41 (2012) 5599-5612.

[22] H. Zhang, M. Chiao, Anti-fouling coatings of poly(dimethylsiloxane) devices for biological and biomedical applications, J. Med. Biol. Eng. 35 (2015) 143-155.

[23] E.P. Magennis, A.L. Hook, M.C. Davies, C. Alexander, P. Williams, M.R. Alexander, Engineering serendipity: High-throughput discovery of materials that resist bacterial attachment, Acta Biomater. 34 (2016) 84-92.

[24] W.H. Kuo, M.J. Wang, H.W. Chien, T.C. Wei, C. Lee, W.B. Tsai, Surface modification with poly(sulfobetaine methacrylate-co-acrylic acid) to reduce fibrinogen adsorption, platelet adhesion, and plasma coagulation, Biomacromolecules 12 (2011) 4348-4356.

[25] A.J. Keefe, N.D. Brault, S. Jiang, Suppressing surface reconstruction of superhydrophobic PDMS using a superhydrophilic zwitterionic polymer, Biomacromolecules 13 (2012) 1683-1687.

[26] S. Chen, L. Li, C. Zhao, J. Zheng, Surface hydration : Principles and applications toward low-fouling / nonfouling biomaterials, Polymer 51 (2010) 5283-5293.

[27] K. Ishihara, Y. Iwasaki, Reduced protein adsorption on novel phospholipid polymers, J. Biomater. Appl. 13

(1998) 111-127.

[28] K. Ishihara, H. Nomura, T. Mihara, K. Kurita, Y. Iwasaki, N. Nakabayashi, Why do phospholipid polymers reduce protein adsorption?, J. Biomed. Mater. Res. 39 (1998) 323-330.

[29] S. Sánchez-Salcedo, A. García, M. Vallet-Regí, Prevention of bacterial adhesion to zwitterionic biocompatible mesoporous glasses, Acta Biomater. 57 (2017) 472-486.

[30] L. Pontiroli, M. Dadkhah, G. Novajra, I. Tcacencu, S. Fiorilli, C. Vitale-Brovarone, An aerosol-spray-assisted approach to produce mesoporous bioactive glass microspheres under mild acidic aqueous conditions, Mater. Lett. 190 (2017) 111-114.

[31] L.T. Zhuravlev, Surface characterization of amorphous silica-a review of work from the former USSR, Colloids Surfaces A Physicochem. Eng. Asp. 74 (1993) 71-90.

[32] R.J. Hunter, Interaction between Colloidal Particles, Zeta Potential Colloid Sci. (1981) 363-369.

[33] A.L.B. Maçon, T.B. Kim, E.M. Valliant, K. Goetschius, R.K. Brow, D.E. Day, A. Hoppe, A.R. Boccaccini, I.Y. Kim, C. Ohtsuki, T. Kokubo, A. Osaka, M. Vallet-Regì, D. Arcos, L. Fraile, A.J. Salinas, A. V. Teixeira, Y. Vueva, R.M. Almeida, M. Miola, C. Vitale-Brovarone, E. Vernè, W. Höland, J.R. Jones, A unified in vitro evaluation for apatite-forming ability of bioactive glasses and their variants, J. Mater. Sci. Mater. Med. 26 (2015) 115-125.

[34] K.K. Kaysinger, W.K. Ramp, Interaction between Colloidal Particles, Zeta Potential Colloid Sci., (1981) 363-369.

[35] D. Lozano, M. Manzano, J.C. Doadrio, A.J. Salinas, M. Vallet-Regí, E. Gómez-Barrena, P. Esbrit, Osteostatin-loaded bioceramics stimulate osteoblastic growth and differentiation, Acta Biomater. 6 (2010) 797-803.

[36] N. Gomez-Cerezo, I. Izquierdo-Barba, D. Arcos, M. Vallet-Regí, Tailoring the biological response of mesoporous bioactive materials, J. Mater. Chem. B (2015) 3810-3819.

[37] M. Guembe, I. Izquierdo-Barba, B. González, M. Vallet-Regí, J. Díez, M. Colilla, D. Pedraza, Mesoporous silica nanoparticles decorated with polycationic dendrimers for infection treatment, Acta Biomater. 68 (2018) 261-271.

[38] N. Encinas, M. Angulo, C. Astorga, I. Izquierdo-barba, M. Vallet-Regí, Mixed-charge pseudo- zwitterionic mesoporous silica nanoparticles with low-fouling and reduced cell uptake properties, Acta Biomater. 84

(2019) 317-327.

[39] F. Hoffmann, M. Cornelius, J. Morell, M. Fröba, Silica-based mesoporous organic-inorganic hybrid materials, Angew. Chemie - Int. Ed. 45 (2006) 3216-3251.

[40] S. Fiorilli, B. Onida, B. Bonelli, E. Garrone, In situ infrared study of SBA-15 functionalized with carboxylic groups incorporated by a Co-condensation route, J. Phys. Chem. B. 109 (2005) 16725-16729.

[41] S. Fiorilli, L. Rivoira, G. Calì, M. Appendini, M.C. Bruzzoniti, M. Coïsson, B. Onida, Iron oxide inside SBA-15 modified with amino groups as reusable adsorbent for highly efficient removal of glyphosate from water, Appl. Surf. Sci. 411 (2017) 457-465.

[42] L. Dalstein, E. Potapova, E. Tyrode, The elusive silica/water interface: Isolated silanols under water as revealed by vibrational sum frequency spectroscopy, Phys. Chem. Chem. Phys. 19 (2017) 10343-10349.

[43] R.K. Iler, The Chemistry of Silica: Solubility, Polymerization, Colloid and Surface Properties and Biochemistry of Silica, Sci.Geol., 35 (1979) 93-94.

[44] I. Izquierdo-Barba, M. Vallet-Regí, Mesoporous bioactive glasses: Relevance of their porous structure compared to that of classical bioglasses, Biomed. Glas. 1 (2015) 140-150.

[45] V. Uskoković, D.P. Uskoković, Nanosized hydroxyapatite and other calcium phosphates: Chemistry of formation and application as drug and gene delivery agents, J. Biomed. Mater. Res. - Part B Appl. Biomater. 96 B (2011) 152-191.

[46] K. Lin, C. Wu, J. Chang, Advances in synthesis of calcium phosphate crystals with controlled size and shape, Acta Biomater. 10 (2014) 4071-4102.

[47] K.K. Kaysinger, W.K. Ramp, Extracellular pH modulates the activity of cultured human osteoblasts, J. Cell. Biochem. 1 (1998) 83-89.

[48] P. Naruphontjirakul, O. Tsigkou, S. Li, A.E. Porter, J.R. Jones, Human mesenchymal stem cells differentiate into an osteogenic lineage in presence of strontium containing bioactive glass nanoparticles, Acta Biomater. 90 (2019) 373-392.

[49] J. Lu, M. Liong, Z. Li, J.I. Zink, F. Tamanoi, Biocompatibility, biodistribution, and drug-delivery efficiency of mesoporous silica nanoparticles for cancer therapy in animals, Small. 6 (2010) 1794-1805.

[50] T.H. Qazi, S. Hafeez, J. Schmidt, G.N. Duda, A.R. Boccaccini, E. Lippens, Comparison of the effects of 45S5 and 1393 bioactive glass microparticles on hMSC behavior, J. Biomed. Mater. Res. A 105 (2017)

2772-2782.

[51] P. Naruphontjirakul, A.E. Porter, J.R. Jones, In vitro osteogenesis by intracellular uptake of strontium containing bioactive glass nanoparticles, Acta Biomater. 66 (2018) 67-80.

[52] I. Izquierdo-Barba, S. Sánchez-Salcedo, M. Colilla, M.J. Feito, C. Ramírez-Santillán, M.T. Portolés, M. Vallet-Regí, Inhibition of bacterial adhesion on biocompatible zwitterionic SBA-15 mesoporous materials, Acta Biomater. 7 (2011) 2977-2985.

[53] J. Zhang, S. Zhao, Y. Zhu, Y. Huang, M. Zhu, C. Tao, C. Zhang, Three-dimensional printing of strontium-containing mesoporous bioactive glass scaffolds for bone regeneration, Acta Biomater. 10 (2014) 2269-2281.